\documentclass[sigconf]{acmart}
\AtBeginDocument{%
  }

\setcopyright{acmlicensed}
\copyrightyear{2018}
\acmYear{2018}
\acmDOI{XXXXXXX.XXXXXXX}
\acmConference[Conference acronym 'XX]{Make sure to enter the correct
  conference title from your rights confirmation email}{June 03--05,
  2018}{Woodstock, NY}
\acmISBN{978-1-4503-XXXX-X/2018/06}

\usepackage{xcolor}
\usepackage{array}
\usepackage{graphicx}
\usepackage{subcaption}
\usepackage{natbib}
\usepackage{mdframed}
\usepackage{hyperref}
\begin{document}

\title{Self-Explanation Tutor for Active Study of CS1 Worked Examples}

\author{Arun-Balajiee Lekshmi-Narayanan}
\email{arl122@pitt.edu}
\orcid{0000-0002-7735-5008}
\affiliation{%
  \institution{University of Pittsburgh}
  \city{Pittsburgh}
  \state{PA}
  \country{USA}
}

\author{Mohammad Hassany}
\email{moh70@pitt.edu}
\orcid{0009-0004-8893-8454}
\affiliation{%
  \institution{University of Pittsburgh}
  \city{Pittsburgh}
  \state{PA}
  \country{USA}
}
\author{Kamil Akhuseyinoglu}
\email{kakhusey@andrew.cmu.edu}
\orcid{0000-0002-7761-9755}
\affiliation{%
  \institution{Carnegie Mellon University}
  \city{Pittsburgh}
  \state{PA}
  \country{USA}
}
\author{Rully Hendrawan}
\email{rah225@pitt.edu}
\orcid{0000-0001-8541-0305}
\affiliation{%
  \institution{University of Pittsburgh}
  \city{Pittsburgh}
  \state{PA}
  \country{USA}
}
\author{Peter Brusilovsky}
\email{peterb@pitt.edu}
\orcid{0000-0002-1902-1464}
\affiliation{%
  \institution{University of Pittsburgh}
  \city{Pittsburgh}
  \state{PA}
  \country{USA}
}








\renewcommand{\shortauthors}{Anon et al.}

\begin{abstract}
Worked examples are an important part of introductory programming, but reading their expert explanations is passive. Self explanation, students explaining the problem and its solution to themselves with subgoal level analysis, converts passive reading into an active study of worked example, yet it is hard to scale because assessing free-text explanations and returning timely feedback has had no easy automated solution. We investigate whether a large language model (LLM) can fill that gap. We build a self-explanation tutor for introductory programming, ESSE, in which students explain lines of worked examples and receive immediate LLM feedback on the correctness and completeness of each explanation, and we pursue two goals. First, we ask whether the LLM judges student explanations well enough to serve as the engine of the tutor; we assess its judgments against two independent human reference standards of different kinds, a single domain expert and a crowd of non-expert raters, each with its own strengths and weaknesses, characterizing both where the LLM is reliable and the systematic tendencies in how it diverges. Second, we ask whether the LLM-based tutoring benefits students; deploying it in an introductory Java course, we find that its feedback leads students to persist and revise rather than abandon a line, that their explanations grow more complete and conceptually richer across attempts, and that students show evidence of learning. These indicate that LLM-based assessment is good enough to power a self-explanation tutor, and that the tutor positively shapes how students study worked examples.
\end{abstract}

\ccsdesc[500]{Applied computing~Interactive learning environments}
\ccsdesc[500]{Human-centered computing~HCI theory, concepts and models}
\ccsdesc[500]{Computing methodologies~Cross-validation}
\ccsdesc[500]{Computing methodologies~Discourse, dialogue and pragmatics}

\keywords{Programming, Self--Explanations, Evaluation, Feedback, LLMs}

\received{}
\received[revised]{}
\received[accepted]{}

\maketitle
\section{Introduction}
In programming education, worked examples present complete solutions whose lines are augmented with expert explanations, helping novices acquire problem-solving schemas before they write code on their own~\cite{linn1992can,brusilovsky2008,hosseini2020improving}. Reading or watching such explanations, however, is \emph{passive}: the ICAP framework~\cite{chi2014icap} argues that learning deepens as activity moves from passive to active and constructive engagement. A way to make worked-example study constructive is \emph{self-explanation}, in which students articulate the purpose and behavior of each step in their own words~\cite{shareghi2013examples,caughey2023investigating}. Self-explanation of code is tied to gains in program comprehension and writing skill~\cite{lopez2008relationships}, yet existing implementations share a practical bottleneck: there is no scalable, reliable, and timely way to grade the free-text explanations students produce and to give them feedback. Prior work has students explain code in their own words but largely stops at eliciting or analyzing those explanations~\cite{leinonen2023comparing}; what has been missing is a tutor that automatically assesses each self-explanation and returns that assessment so the student can revise.
 
The bottleneck is challenging because correct student explanations vary enormously in wording. Approaches that score an explanation by its \emph{semantic similarity} to a single expert reference penalize correct answers phrased unlike the expert~\cite{rus2013semilar,oli2024can}. Recent work shows that prompting an LLM to act as a judge of student explanations can match or exceed fine-tuned similarity models~\cite{pmlr-v257-oli24a}, opening the door to feedback that does not require an exact reference answer. This makes it feasible to build an interface around the active self-explanation task-one that scores each explanation and returns the score to the student in real time-and to ask what such an interface affords once students use it.
 
This paper presents a pilot study we ran with a user interface in a real introductory Java course. The tool is itself a pedagogical intervention: it converts the passive act of reading expert explanations into an active writing task and surrounds that task with layered, immediate feedback designed to keep students productively engaged rather than stuck. For each selected line it returns a judgment of \emph{correctness} (binary) and \emph{completeness} (continuous) together with counts of the concepts present and absent. We study the interface we built around two research questions:
 
\begin{itemize}
  \item[\textbf{RQ1}] To what extent can an LLM judge the correctness and completeness of student self-explanations-well enough to support a tutor-style interaction that assesses each explanation and returns that assessment to the student?
  \item[\textbf{RQ2}] What impact does the tutor have on students' self-explanation behavior as they study worked examples-in particular, do their explanations improve as they attempt more?
\end{itemize}
 
\noindent\textbf{Contributions.} (1) We present an interactive self-explanation tool for worked examples that delivers immediate, concept-level LLM feedback, and report a pilot of it in CS1. (2) On live CS1 data we characterize \emph{how} the LLM and a human disagree-high raw agreement but a systematic under-scoring bias-and show why a single agreement statistic is misleading under skewed label distributions. (3) We corroborate this against a second, independent human standard-a reliability-filtered crowd-using prevalence-robust statistics suited to skewed labels. (4) We show that explanation \emph{completeness}, and specifically its conceptual content rather than its length, is the signal that tracks student revision and improvement.

  


\section{Related Work}
 
\subsection{Self-Explanation and Active Study of Worked Examples} 
Self-explanation is a robust learning strategy across domains~\cite{chi2018learning}, and within the ICAP framework~\cite{chi2014icap} it is a constructive activity that outperforms passive review. In programming specifically, prompting students to explain code has been integrated into tutors for SQL~\cite{shareghi2013examples} and Python~\cite{caughey2023investigating}, both reporting comprehension gains, and self-explanation has been contrasted with reading worked examples and code comments~\cite{oli2023reading}. Example systems such as PCEX~\cite{hosseini2018pcex} let students step through worked examples line by line, but reading expert explanations remains passive. 

While computer science was one of the first domains where the power of self-explanations was explored, active research of using self-explanations in this field started very recently. So far, self-explanations in this area explored several ideas - explaining programming concepts when solving a problem ~\cite{Fabic2019}, explaining fragments of code examples~\cite{Fabic2019,tamang2021comparative,rus2021prompting}, and explaining preparatory materials~\cite{sibia_flipped_db_course_2024,wen_se_timing_2025,wen_se_llm_2025}.  These experiments brough positive results.  For example, the study~\cite{tamang2021comparative}, reported that the Socratic method of guided (scaffolded) self-explanation is more effective than free self-explanation in teaching code comprehension skills among students enrolled in an introductory computer science course. A classroom study exploring self-explanations of educational videos in the context of a database course has found correlations between self-explanations and student understanding, as well as classroom performance~\cite{sibia_flipped_db_course_2024}.

The main technical challenge in deploying self-explanation approaches in computer science courses is developing reliable approaches for assessing self explanations. Scalable tools for evaluating student explanations in the context of learning from worked examples used a simple menu-based approach~\cite{conati2000further,Fabic2019}. However, research demonstrated that modern semantic similarity approaches could be used to support an efficient dialog with students in computer science education~\cite{rus2013recent,rus2021prompting}. In our work, we bring together previous experience in supporting self-explanation in the domain of computer science education~\cite{bielaczyc1995training,Fabic2019} with automatic assessment and a scaffolded dialog based on recent semantic similarity research.
 
\subsection{Automated Assessment of Explanations and LLM-as-a-Judge}
Automatically evaluating free-form student responses has a long history in intelligent tutoring, beginning with semantic-similarity comparison against an instructor's model answer. Latent Semantic Analysis, introduced in AutoTutor, was the first such technique~\cite{Graesser01082000,wild2005parameters}, and the SEMILAR toolkit later made semantic similarity a standard tool for scoring student responses~\cite{rus2013semilar,rus2013recent,banjade2016evaluation}. Deep-learning successors replaced hand-built similarity with sentence
embeddings, for example a transformer extension of SEMILAR built on
SentenceBERT~\cite{chapagain2022automated,reimers2019sentence}. All of these methods share a structural limitation: because they score an explanation by its proximity to an expert reference, they reward students who echo the expert's wording over those who express the same idea differently~\cite{lekshmi2024evaluating}, and they reveal little about \emph{which} concepts an explanation is missing-the information most useful for feedback. The same reference-matching bias affects general-purpose overlap and embedding metrics such as BLEU, ROUGE, and BERTScore~\cite{Papineni2002BleuAM,lin2004rouge,zhang2019bertscore}.
 
Large language models let a grader assess an explanation directly, reasoning about the concepts it should contain rather than matching a reference. Prompted with few-shot examples and chain-of-thought reasoning, an LLM can judge the correctness of a line-level explanation as well as or better than fine-tuned similarity models~\cite{pmlr-v257-oli24a,oli2024can}, and this \emph{LLM-as-a-judge} paradigm has grown rapidly~\cite{li2025generation}. 

 
Closest to our work are recent systems that use LLMs to assess code
self-explanations. An open-source LLM was fine-tuned with preference optimization and embedded in a dialogue-based tutor in which students make repeated attempts to explain a line of code~\cite{pmlr-v257-oli24a,oli2024exploring}. Explanation correctness has also been assessed indirectly by submitting a student's explanation to an LLM as a specification, generating code from it, and checking whether that code behaves like the original~\cite{denny2024explaining}. We take a more direct route: we prompt an LLM to judge each line-level explanation's correctness and completeness and to report the concepts it covers, without fine-tuning and without matching an expert reference. We then benchmark this judge against a semantic-similarity baseline on a public dataset~\cite{chapagain2025selfcode}, audit its judgments against human raters on live data, and use its concept-level output to drive feedback.

\section{ESSE: An LLM-Powered Self-Explanation Tutor}
 
\begin{figure}[t!]
  \centering
  \includegraphics[width=\columnwidth]{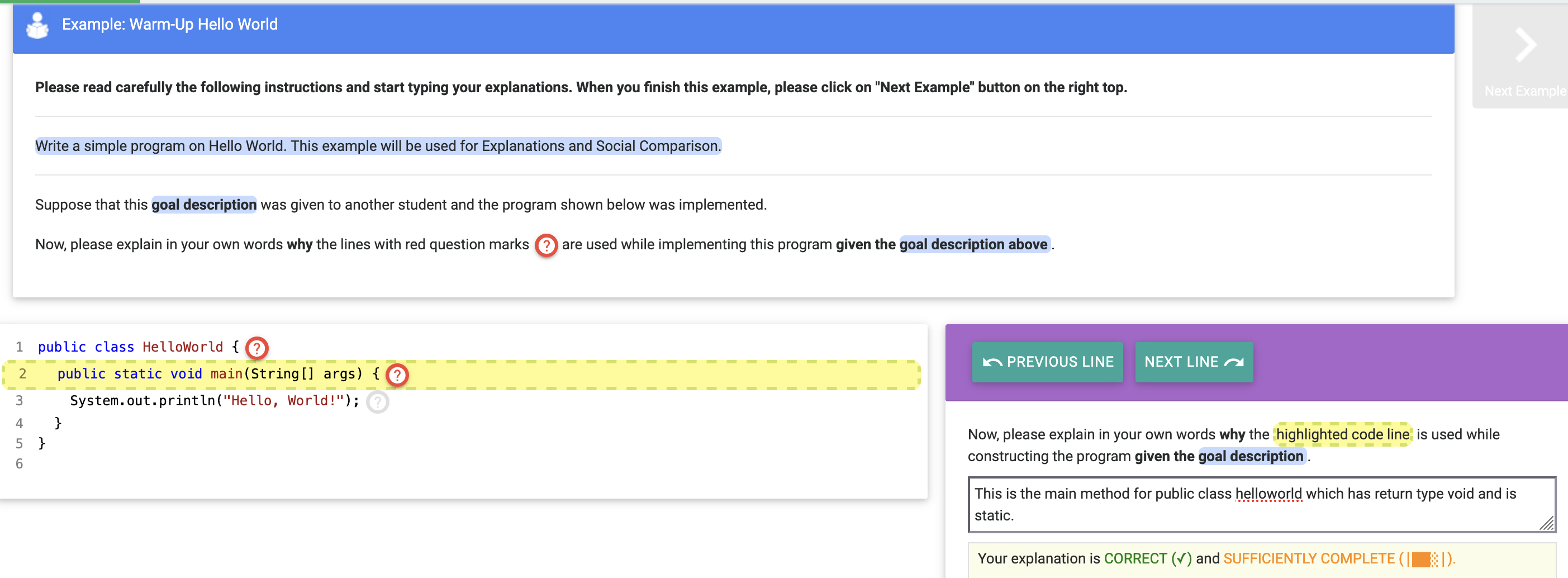}
  \caption{The ESSE self-explanation tutor. Students explain a selected line of a worked example in their own words and receive immediate LLM feedback on the \emph{correctness} and \emph{completeness} of the explanation.}
  \label{fig:system}
\end{figure}

We developed a self-explanaton tutor ESSE (\emph{Example Study with
Self-Explanations}) (ESSE), a new type of standalone interactive ``smart content'' or embedded in a any personalized practice portal such as Mastery Grids~\cite{loboda2014mastery}. ESSE replaces the passive ``read the expert explanation'' step of a worked-example viewer with an active ``write your own explanation'' step, then returns immediate feedback (Figure~\ref{fig:system}). The design goal is pedagogical: move the student from passive reading toward constructive articulation, by the ICAP framework~\cite{chi2014icap}, while keeping the cost of getting unstuck low. In placing an LLM at the center of assessment and feedback, ESSE follows recent generative intelligent tutoring systems~\cite{liu2024advancing}, but targets line-level self-explanation of worked examples.
 
\subsection{The Self-Explanation Activity}
A student works through a worked example one line at a time. For each
highlighted line, ESSE asks the student to explain, in their own words, why that line is used while constructing the program given the goal description, and to submit the explanation. The first attempt is unscaffolded. On submission, ESSE evaluates the explanation and returns feedback; the student may revise and resubmit as often as they like, and the control to advance to the next line becomes available once the explanation is judged correct. The tutor's targets are explanations that are both \emph{correct} and \emph{complete}: a correct explanation covers at least the most important concepts and focuses on the behavior of the line, and a complete explanation covers all and only the concepts expected for that line.
 
\subsection{LLM Assessment of Explanations}
\begin{figure}[t!]
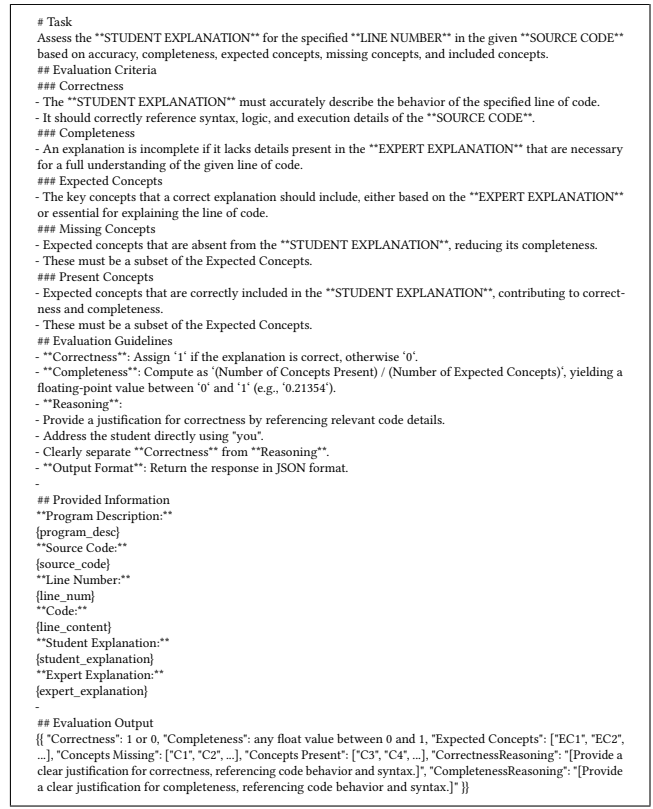

    \centering
    \tiny
    \begin{mdframed}[align=center]
    \# Task  

Assess the **STUDENT EXPLANATION** for the specified **LINE NUMBER** in the given **SOURCE CODE** based on accuracy, completeness, expected concepts, missing concepts, and included concepts.

\#\# Evaluation Criteria  

\#\#\# Correctness  

- The **STUDENT EXPLANATION** must accurately describe the behavior of the specified line of code.  

- It should correctly reference syntax, logic, and execution details of the **SOURCE CODE**.

\#\#\# Completeness  

- An explanation is incomplete if it lacks details present in the **EXPERT EXPLANATION** that are necessary for a full understanding of the given line of code.  

\#\#\# Expected Concepts  

- The key concepts that a correct explanation should include, either based on the **EXPERT EXPLANATION** or essential for explaining the line of code.

\#\#\# Missing Concepts  

- Expected concepts that are absent from the **STUDENT EXPLANATION**, reducing its completeness.  

- These must be a subset of the Expected Concepts.  

\#\#\# Present Concepts  

- Expected concepts that are correctly included in the **STUDENT EXPLANATION**, contributing to correctness and completeness.  

- These must be a subset of the Expected Concepts.  

\#\# Evaluation Guidelines  

- **Correctness**: Assign `1` if the explanation is correct, otherwise `0`.  

- **Completeness**: Compute as `(Number of Concepts Present) / (Number of Expected Concepts)`, yielding a floating-point value between `0` and `1` (e.g., `0.21354`).  

- **Reasoning**:  

  - Provide a justification for correctness by referencing relevant code details.
  
  - Address the student directly using "you".  
  
  - Clearly separate **Correctness** from **Reasoning**.  
  
- **Output Format**: Return the response in JSON format.  

-

\#\# Provided Information  

**Program Description:**  

\{program\_desc\}  

**Source Code:**  

\{source\_code\}  

**Line Number:**  

\{line\_num\}  

**Code:**  

\{line\_content\}  

**Student Explanation:**  

\{student\_explanation\}  

**Expert Explanation:**  

\{expert\_explanation\}  

-

\#\# Evaluation Output  

\{\{
  "Correctness": 1 or 0,
  "Completeness": any float value between 0 and 1,
  "Expected Concepts": ["EC1", "EC2", ...],
  "Concepts Missing": ["C1", "C2", ...],
  "Concepts Present": ["C3", "C4", ...],
  "CorrectnessReasoning": "[Provide a clear justification for correctness, referencing code behavior and syntax.]",
  "CompletenessReasoning": "[Provide a clear justification for completeness, referencing code behavior and syntax.]"
\}\}

    \end{mdframed}
        \caption{The evaluation prompt used by ESSE. A single prompt produces all three quantities from the student explanation and its context.}
        \label{fig:eval-prompt}
\end{figure}

ESSE scores each submitted explanation on two dimensions: \emph{correctness}, a binary judgment of whether the explanation captures the line's behavior, and \emph{completeness}, a continuous score in $[0,1]$, with an explanation deemed ``sufficiently complete'' at $\ge 0.5$. A single prompt to GPT-4o mini performs the assessment: given the problem statement, the worked example, the target line, and an expert explanation as context, the model returns the correctness label, the completeness score, and the numbers of expected concepts that are present in and absent from the student's explanation (Figure~\ref{fig:eval-prompt}). The expert explanation is supplied only as context for judgment, not as an answer key to match, so the assessment requires no exact reference answer at scoring time.
 
\subsection{Layered Feedback}
\label{sec:feedback}
Feedback is delivered as four levels of increasing scaffolding, unlocked
progressively: the student reaches the next level only after engaging the
current one and revising their explanation.
\begin{itemize}
  \item \textbf{Flags} - shown after every submission: a correctness indicator
  (green/red, labelled correct or incorrect) and a completeness indicator
  (a bar with a green, yellow ``sufficiently complete,'' or red state).
  \item \textbf{Samples} - peer-written explanations from a prior
  dataset~\cite{chapagain2025selfcode} that a student can browse for extra detail,
  including examples attributed to more and less experienced students.
  \item \textbf{Feedback} - an LLM-generated message naming the concepts the
  explanation is missing and how to improve it, which the student can rate
  (like/dislike).
  \item \textbf{Answer} - an expert explanation of the line.
\end{itemize}
\noindent Flags appear immediately, but Samples, Feedback, and the Answer are gated: the Feedback and Answer options stay locked until the student has consulted the earlier levels and resubmitted, so the tutor withholds the solution until students have attempted revision themselves. Advancing to the next line requires a correct explanation; completeness is instead encouraged through the flags.

\section{Study Design and Data}
 
We draw on two datasets (Table~\ref{tab:datasets}): a deployed CS1 pilot and an independent crowd re-rating of its explanations.
 
\paragraph{Deployed CS1 study (RQ1, RQ2).}
As a pilot of ESSE, we deployed it as a recitation activity in an
introductory Java course (CS1). All $N=8$ enrolled students participated under consent. Following the design of prior studies in this line of
work~\cite{hosseini2018pcex}, students first took a pretest assessing prior Java knowledge; they then worked through ESSE, explaining 30 lines of code across four intermediate worked examples (conditions, loops, objects); and finally took an isomorphic posttest. Participation earned extra credit and was not graded on correctness, to encourage genuine attempts rather than answer-matching. ESSE logged every interaction: students produced 409 attempts, of which 407 received feedback. At the end of the session, students completed a short survey on their perceptions of the tool and its feedback. A research expert annotated each live explanation--attempt for correctness and completeness, providing the human standard for RQ1. The change and agreement analyses below use the 358 feedback events that pair an attempt with its predecessor.
 
\paragraph{Independent crowd study (RQ1).}
To obtain a second, independent set of human labels, we recruited a crowd on Amazon Mechanical Turk to re-rate live-study explanations: 124 workers produced 1{,}696 ratings over 216 explanations on the same correctness and completeness rubric, with uneven coverage across explanations. Aggregated non-expert judgments can approximate expert labels for many annotation tasks~\cite{snow2008cheap}. We filter unreliable workers by chance-corrected agreement against the majority ($\kappa$-vs-majority), since raw agreement passes ``base-rate clickers'' who always select the majority label~\cite{raykar2010learning,hovy2013learning}, and aggregate the reliable raters into a single ``community-wisdom'' label per explanation.
 
\begin{table}[t]
\centering
\small
\setlength{\tabcolsep}{4pt}
\caption{The two datasets and the research questions they support.}
\label{tab:datasets}
\begin{tabular}{@{}llll@{}}
\toprule
Dataset & Source & Size & Used for \\
\midrule
CS1 study & Deployment & $N{=}8$; 409 attempts        & RQ1, RQ2 \\
Crowd     & MTurk      & 124 raters; 1,696 ratings    & RQ1 \\
\bottomrule
\end{tabular}
\end{table}
 
\paragraph{Analysis and notation.}
For behavioral models (RQ2) we fit mixed-effects models with a random intercept for student (\texttt{user\_id}) and report Satterthwaite-approximated significance. For agreement (RQ1) we report precision, recall, and F1 with McNemar's test for the binary correctness dimension, and point-biserial correlation and AUC for the continuous-versus-binary completeness comparison; crowd worker reliability is scored by chance-corrected agreement against the majority ($\kappa$-vs-majority). Throughout, we denote significance as $p<.05$ ($*$), $p<.01$ ($**$), and $p<.001$ ($***$), and follow APA convention in dropping the leading zero from coefficients bounded by one.
 
\section{Evaluation}
\subsection{RQ1: Building an Interface with Reliable Automated Feedback}
 
The interface needs a feedback engine whose judgments students can act on, so
we ask how well the LLM's correctness and completeness scores agree with human judgment. We have two independent sets of human labels-a single expert
(\S\ref{sec:rq1-agreement}) and a crowd of non-expert raters
(\S\ref{sec:rq1-crowd})-and we report how well the LLM does against each as a standard in turn.
 
\subsubsection{LLM vs.\ the Expert}
\label{sec:rq1-agreement}
 
\paragraph{Correctness.}
On the deployed study's 358 rated attempts, treating the expert's ``correct'' label as the positive class, the LLM reaches precision $.95$ and recall $.87$ (F1~$=.91$; accuracy $83.8\%$). The gap between precision and recall is the telling part: the LLM misses correct explanations far more often than it accepts incorrect ones-42 false negatives against 16 false positives (McNemar $p=.001$), so it systematically \emph{under-scores} correctness. Chance-corrected agreement is correspondingly only fair ($\kappa=.32$) under the skewed label distribution.
 
\paragraph{Completeness.}
Because the LLM emits completeness as a continuous score while the expert
records a binary judgment, we treat the expert label as the outcome and ask how well the score separates it: the LLM's completeness score is a significant predictor (point-biserial $r=.39$, $p<.001$; AUC~$=.67$) but poorly calibrated in absolute terms (mean absolute error $=.45$ against the binary label), tracking human judgment while falling short of a clean separation.
 
\subsubsection{LLM vs.\ the Crowd}
\label{sec:rq1-crowd}
The crowd gives a second, independent human standard. Because each explanation was rated by several workers, we aggregate their ratings into one ``community-wisdom'' label per explanation and dimension. Raw agreement would let base-rate ``clickers''-who select the majority label on nearly every item under the skewed distribution-count as reliable, so we score each worker by chance-corrected agreement with the crowd majority ($\kappa$-vs-majority), keeping 58 reliable raters (9 unreliable; the rest rated too few items to assess), and form each crowd label from a reliability-weighted majority vote.
 
\paragraph{Correctness.}
Against the crowd standard (216 explanations), the LLM reaches precision $.90$ and recall $.95$ (F1~$=.93$). Its few errors lean toward false positives (19) over false negatives (9)-the reverse of the expert comparison, where the LLM withheld credit; against the more lenient crowd it grants it.
 
\paragraph{Completeness.}
Completeness agreement is weaker, consistent with the crowd itself agreeing less on this dimension. The LLM's continuous completeness score orders the crowd judgment moderately (AUC~$=.76$) but is poorly calibrated in absolute terms (mean absolute error $=.42$ against the crowd's fraction-complete).
 
\subsection{RQ2: What Students Engage With, and What Impacts Learning}
 
\subsubsection{Intrinsic: Change in Explanation Quality}
 
The impact question for RQ2 is whether the feedback moves students toward better explanations. A good explanation covers the concepts a line of code requires, so we ask students whether they revise their response when they receive feedback. We also ask whether their explanations then become more \emph{complete}-by adding the \emph{right concepts} rather than simply more words.
 
\paragraph{Students revise when the feedback flags a problem.}
Students made $1.66$ attempts per explanation on average (median $1$, max $7$), and revision does not taper off with effort: $38\%$ of explanations receive a second attempt ($83$ of $216$) and $37\%$ of those a third ($31$ of $83$), with the per-step continuation rate holding or rising among the smaller group who go further. Whether they attempt again is strongly conditioned on the feedback they receive (Table~\ref{tab:persistence}): they revise $63\%$ of the time after an
\emph{incorrect} verdict but only $35\%$ after a correct one, and their
willingness to revise climbs steadily as completeness falls---from $0\%$ when the explanation is already near-complete to about two-thirds when it is least complete. Students treat the feedback as actionable, revising precisely when it signals a gap and stopping when the explanation is good.

\begin{table}[t]
\centering
\small
\caption{Share of attempts followed by another attempt (``revise''), by the
correctness and completeness of the current attempt. Students revise more when
the feedback flags a problem.}
\label{tab:persistence}
\begin{tabular}{@{}lrr@{}}
\toprule
Current attempt & Attempts & Revise \\
\midrule
\multicolumn{3}{@{}l}{\textit{by correctness}}\\
\quad Incorrect           & 62  & 63\% \\
\quad Correct             & 296 & 35\% \\[2pt]
\multicolumn{3}{@{}l}{\textit{by completeness}}\\
\quad $\le 25\%$          & 14  & 64\% \\
\quad 25--50\%            & 271 & 42\% \\
\quad 50--75\%            & 66  & 30\% \\
\quad $>75\%$             & 7   & \phantom{0}0\% \\
\bottomrule
\end{tabular}
\end{table}
 
\paragraph{Revised explanations become more complete, by adding concepts.}
Across successive attempts, completeness scores rise substantially (Wilcoxon
$r=.84$, $p<.001$). The gain reflects \emph{what} students add, not merely how
much: completeness is \emph{negatively} correlated with the change in word count
(Spearman $r=-.49$, $p<.001$), and in a mixed-effects model predicting
completeness from the number of distinct concepts and the word count (random
intercept per student), conceptual content dominates volume
($\beta_{\text{concept}}=.50^{***}$ vs.\ $\beta_{\text{volume}}=.33^{***}$).
Because completeness is defined by concept coverage, students improve by
supplying the concepts a line requires rather than padding their prose-the
behavior the feedback is meant to induce.
 
\subsubsection{Extrinsic: Learning Gain from Pretest to Posttest}
 
\paragraph{Learning gain by prior knowledge.}
Only 3 of the 8 students completed the posttest; the other 5 did not and are
excluded rather than scored zero. Among the three completers, normalized gains
were $.67$, $-.25$, and $.78$, and the highest gain came from the student with
the lowest pretest (1 of 9). The pilot is too small for this to be more than
suggestive.
 
\subsubsection{Student Perceptions}
 
\begin{figure}[t]
    \centering
    \includegraphics[width=\columnwidth]{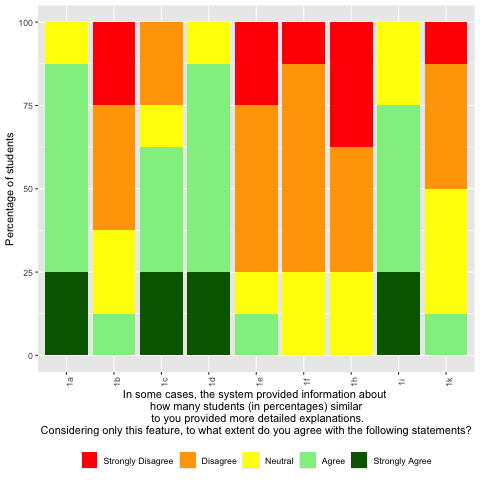}
    \caption{Students' attitudes toward using the ESSE system.}
    \label{fig:survey}
\end{figure}

The survey corroborates the behavioral findings (Figure~\ref{fig:survey},
Table~\ref{tab:survey}). First, students rated the feedback well overall: six of
eight disagreed that it was irrelevant (1e) or incorrect or misleading (1f), and
six of eight disagreed that they would use the system \emph{without} feedback
(1h)-they valued it. Two items line up directly with the objective results
above: five of eight agreed the feedback helped them write more (1c), matching
the measured growth in explanation length, and seven of eight agreed it helped
them recall the concepts to include (1d), matching the concept-driven
completeness gains. Students were also clear on the activity, understanding what
a correct explanation should be (1a) and the goal of the task (1i); the least
decisive responses concerned the definition of a \emph{complete} explanation
(1b) and knowing what to do next from a given feedback message (1k).

\begin{table}[t]
\centering
\footnotesize
\setlength{\tabcolsep}{3pt}
\renewcommand{\arraystretch}{1.1}
\caption{Survey responses ($N=8$): counts agreeing (A; strongly agree $+$ agree),
neutral (N), and disagreeing (D; disagree $+$ strongly disagree).
$^\dagger$Items phrased so that \emph{disagreement} is the favorable response.}
\label{tab:survey}
\begin{tabular}{@{}llccc@{}}
\toprule
 & Statement (abbreviated) & A & N & D \\
\midrule
1a            & Understood ``correct'' definition          & 7 & 1 & 0 \\
1b$^\dagger$  & Did not understand ``complete'' definition & 1 & 2 & 5 \\
1c            & Feedback helped write longer explanations  & 5 & 1 & 2 \\
1d            & Feedback helped recall key concepts        & 7 & 1 & 0 \\
1e$^\dagger$  & Feedback was irrelevant                    & 1 & 1 & 6 \\
1f$^\dagger$  & Feedback was incorrect or misleading       & 0 & 2 & 6 \\
1h$^\dagger$  & Would use the system without feedback      & 0 & 2 & 6 \\
1i            & Feedback helped understand the task goal   & 6 & 2 & 0 \\
1k$^\dagger$  & Unclear what to do next from feedback      & 1 & 3 & 4 \\
\bottomrule
\end{tabular}
\end{table}

\section{Discussion}
 
\paragraph{Correctness is reliable; its residual error is standard-dependent.}
The LLM's correctness judgments agree strongly with both human standards
(F1~$=.91$ against the expert, $.93$ against the crowd), which is what a tutor needs to act on them. The small residual disagreement points in opposite directions depending on the standard: against the stricter expert the LLM \emph{withholds} credit (more false negatives), while against the more lenient crowd it \emph{grants} it (more false positives), so no single bias dominates. In a formative self-explanation setting the under-scoring direction is the safer one-an unearned ``incorrect'' nudges a student to revise, and we observe that students do revise rather than quit (RQ2), whereas an unearned ``correct'' would let a misconception stand.
 
\paragraph{Choose metrics that fit skewed labels and two-part judgments.}
Our RQ1 analysis is a methodological caution. With the skewed label distributions typical of ``mostly correct'' student work, a single agreement number-raw agreement or Cohen's $\kappa$-is misleading; precision and recall separate the kinds of error that matter for feedback. For a continuous completeness score, ordering (AUC) and absolute calibration (MAE) can diverge sharply-ours ranks completeness moderately well yet is poorly calibrated-so both should be reported rather than a single figure. Filtering raters by chance-corrected reliability before aggregating a crowd standard is a further safeguard against base-rate noise.
 
\paragraph{Make completeness, and its concepts, the actionable target.} Since completeness-driven by conceptual content-is what tracks revision and improvement, feedback should foreground \emph{which concepts are missing} rather than reward length. Our concept-level feedback is one realization of this; the result argues for designing feedback around concepts in general.
 
\paragraph{Design implications for self-explanation tools.}
As a pilot, the study also yields concrete design guidance. The faded
scaffolding-withholding the expert answer until later levels-coexisted with productive persistence: students revised rather than quit (RQ2), suggesting the fade from flags to samples to feedback to answer is worth retaining rather than short-circuiting with an early reveal. And because conceptual completeness is the signal that moves learning, the interface should surface missing concepts as the primary feedback message rather than emphasizing a raw score or word count.
 
\section{Limitations}
The deployment is a single CS1 section with $N=8$ students-a pilot-so RQ2
results, especially the learning-gain pattern, are suggestive rather than
confirmatory; only 3 of 8 students have valid posttest data (a posttest of 0 indicates non-completion), and the perception-survey data is preliminary and small. The human standard in RQ1 is one expert; we add a second, independent crowd standard (also RQ1) but cannot rule out a shared blind spot. We also evaluate a single LLM grader under a single prompt configuration.

\section{Future Work}
This pilot opens several directions. First, we plan to move from pilot to
confirmatory evidence with a larger, semester-long study across
multiple CS1 sections, powered for the pretest--posttest learning-gain analysis that $N=8$ could only suggest. Second, ESSE is designed as smart content for a personalized practice portal, and a natural next step is to integrate it into Mastery Grids~\cite{loboda2014mastery} so that the concepts ESSE detects as \emph{present}, \emph{missing}, and \emph{expected} feed the same concept-level open learner model that drives the portal's other activities-letting self-explanation evidence update the knowledge model alongside problem solving. Third, those concept-level signals could in turn drive adaptive support: recommending the next worked example, or surfacing targeted feedback, based on the concepts a learner repeatedly omits-an approach our group has found especially valuable for lower-prior-knowledge students~\cite{barria2021explainable}. Fourth, to remove the single-rater coverage gap, we are running a top-up crowdsourcing round on the 72 under-covered explanations (96 additional ratings) to reach at least three independent crowd raters per explanation; this lets us exclude the researcher from the crowd entirely and treat the expert labels as a clean held-out ground truth rather than a participant in the consensus. Finally, on the evaluation side, we will address the grader's under-scoring bias through calibration and prompt refinement, compare alternative (including open-weight) LLM judges, measure the same-prompt reliability of the judge, and collect graded rather than binary human completeness to enable a richer continuous-to-continuous validation.

\section{Conclusion}
We deployed an LLM-powered self-explanation tutor in CS1 and asked whether its automated feedback is trustworthy enough to drive the tutor and whether it helps students. On live data the LLM's correctness judgments agree strongly with two independent human standards-a single expert and a reliability-filtered crowd-though completeness remains the harder dimension to score. And the feedback changes behavior in the right way: students revise when it flags a problem and stop when their explanation is good, and their explanations grow more complete by adding the concepts a line requires rather than more words. Together these results give cautiously positive evidence for LLM-assisted self-explanation in introductory programming, while making its evaluative blind spots explicit.

\begin{acks}
Thanks to Dr. Xiang Lorraine Li for contributions to this work. This work was supported partially by the Grad Student Provost Fellowship of the Intelligent Systems Program at the University of Pittsburgh and the NSF Award Number 1822752.
\end{acks}

\bibliographystyle{ACM-Reference-Format}
\bibliography{references}



\end{document}